\documentclass[prb,twocolumn,superscriptaddress,showpacs,amsmath,amssymb]{revtex4}

\usepackage{graphicx}
\usepackage{dcolumn}
\usepackage{bm}

\begin{document}


\title{Magnetic anisotropy in thin films of Prussian blue analogues} 

\author{D. M. Pajerowski}
\affiliation{Department of Physics and the National High Magnetic Field Laboratory, University of Florida, 
Gainesville, FL 32611-8440, USA}
\author{J. E. Gardner}
\author{M. J. Andrus}
\affiliation{Department of Chemistry, University of Florida, Gainesville, FL 32611-7200, USA}
\author{S. Datta}
\affiliation{National High Magnetic Field Laboratory and Department of Physics, Florida State 
University, Tallahassee, FL 32310-3706, USA} 
\author{A.~Gomez}
\author{S. W. Kycia}
\affiliation{Department of Physics, University of Guelph, Guelph, ON N1G 2W1, Canada}
\author{S. Hill}
\affiliation{National High Magnetic Field Laboratory and Department of Physics, Florida State 
University, Tallahassee, FL 32310-3706, USA} 
\author{D. R. Talham}
\affiliation{Department of Chemistry, University of Florida, Gainesville, FL 32611-7200, USA}
\author{M. W. Meisel}
\affiliation{Department of Physics and the National High Magnetic Field Laboratory, University of Florida, 
Gainesville, FL 32611-8440, USA}

\date{\today}

\begin{abstract}
The magnetic anisotropy of thin ($\sim 200$~nm) and thick ($\sim 2$~$\mu$m) 
films and of polycrystalline (diameters $\sim 60$~nm)  
powders of the Prussian blue analogue 
Rb$_{0.7}$Ni$_{4.0}$[Cr(CN)$_6$]$_{2.9} \cdot n$H$_2$O, a ferromagnetic 
material with $T_c \sim 70$~K, have been investigated by 
magnetization, ESR at 50 GHz and 116 GHz, and variable-temperature 
x-ray diffraction (XRD).  
The origin of the anisotropic magnetic response cannot be attributed to 
the direct influence of the solid support, but the film growth protocol  
that preserves an organized two-dimensional film is important.  In addition, 
the anisotropy does not arise from an anisotropic g-tensor nor from 
magneto-lattice variations above and below $T_c$. 
By considering effects due to magnetic domains and demagnetization factors, 
the analysis provides reasonable descriptions of the low and high field data, 
thereby identifying the origin of the magnetic anisotropy.
\end{abstract}

\maketitle

\section{Introduction}
There is an increasing demand for novel architectures that afford the possibility
of spin polarized electron transport, a field known as spintronics.\cite{Wolf}  
A key element involves control of the magnetic anisotropy in ferromagnetic 
films and nanostructures.  Accordingly, the ability to manipulate the
underlying magnetic states of the spin polarizers is desirable.    
In addition to traditional solid-state materials, molecule-based magnetic systems are being
investigated.\cite{Bogani,Camarero,Moritomo}    
The discovery of large and persistent photoinduced changes in the magnetization in 
some examples of cyanometallate coordination polymers makes them attractive materials 
to consider.\cite{Sato,Paj-jacs-com}  

Herein, 
studies of the magnetic anisotropy of 
thin and thick films along with standard powder-like samples  
of bimetallic Prussian blue analogues, 
A$_j$M$^{\prime}_k$[M(CN)$_6$]$_{\ell} \cdot n$H$_2$O, where A is
an alkali ion and M$^{\prime}$ and M are transition metal ions,\cite{Dunbar,Verdaguer} 
are reported. Previously, the anisotropic 
response of the persistent photoinduced magnetism of thin films of 
Rb-Co-Fe (referring to A-M$^{\prime}$-M) Prussian blue analogues was discovered\cite{Park1} and 
subsequently studied 
systematically.\cite{Park2,Frye,Park-thesis,Frye-thesis,Gardner-thesis,Pajerowski-thesis}  
The motivation to understand the origins of this anisotropic phenomenon 
is amplified by the ability to control the 
magnetization of Prussian blue analogues 
by photo-irradiation\cite{Sato,Bleuzen1,Paj-jacs-com} or 
pressure.\cite{Zentkova}  However, the magnetic 
response of the photo-controllable A-Co-Fe system is complicated by the multiple stable oxidation 
states of the Co and Fe ions and by orbital angular momentum 
contributions.  Consequently, the Rb-Ni-Cr Prussian blue analogue, a ferromagnet 
system possessing a spectrum of long-range ordering temperatures, 
$T_c \sim 60 - 90$~K, depending on stoichiometry,\cite{Verdaguer} 
was chosen as the centerpiece for the present work because the magnetic and physical properties 
of this system are robust and the ions have stable oxidation states that possess no first-order 
angular momentum.

Finally, it is important to stress the significance of our findings.  Although the study of the 
magnetism of solid-state films is a mature field, the extensions to molecule-based magnetism are 
just beginning to emerge.  For example, with the drive to develop new devices, 
applications with single crystals 
are being explored.\cite{Schmidt}  However, the exploitation of molecule-based magnetic films 
may be more attractive for industrial fabrication, and devices based on 
metal-phthalocyanines\cite{Heutz} and 
metal[TCNE:tetracyanethylene]$_x$\cite{Yoo} are two 
examples of work in this direction.  In our work, the origins of the  
magnetic anisotropy in films of Prussian blue analogues will be linked to demagnetization effects after we have 
systematically eliminated all other plausible explanations, some of which are not issues in 
traditional solid-state magnetic films.  As a result, our results provide 
a foundation from which the magnetism in films of Prussian blue analogues 
may be understood and employed in new devices.

\begin{figure}[ht]
\includegraphics[width=3.375in]{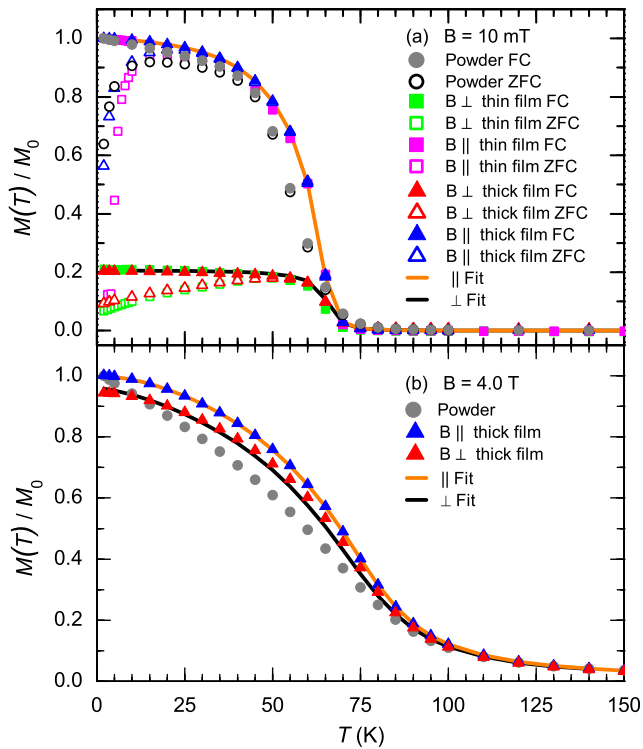}
\caption{(Color online) The temperature dependences of the zero-field-cooled (ZFC) 
and field-cooled (FC) magnetizations, $M(T)$, 
normalized to the FC values at $T = 2$ K, $M_0$, are shown 
for (a) low, $B = 10$ mT, and (b) high, $B = 4$ T,  
applied magnetic fields.  For clarity, the data for the thin film are not 
shown in (b).  The anisotropic 
response for $B$ applied parallel ($\parallel$) or perpendicular ($\perp$) to 
the films is strikingly similar for both thin and thick films.  
The field-induced shift of $T_c$ from 
$\sim 70$ K to $\sim 100$ K is observable.  
For each panel, the solid lines are the results of analysis using 
demagnetization factors (see text).}
\end{figure}

\section{Experimental Details}
The synthesis of the powder samples followed established 
protocols,\cite{Gardner-thesis} while the films were generated using  
sequential adsorption methods\cite{Culp} that are detailed elsewhere.\cite{Gardner-thesis} 
Briefly stated, the film synthesis consists of using a solid support, 
such as Melinex 535, and immersing it in an aqueous solution of Ni$^{2+}$ ions 
and then in another aqueous solution of Cr(CN)$^{3-}_{6}$ containing Rb$^+$ ions.  
After each immersion step, washing with water is essential 
to remove the excess ions, and the process can be iterated multiple cycles to 
yield films of varying thicknesses and morphologies.  For this work, 
two films, one of 40~cycles and the other of 400~cycles, are reported.  
Whereas the powder samples consisted of small polycrystals with diameters of 
$\sim 60$~nm, which are magnetically in the ``bulk'' limit,\cite{dpaj-nano}  
the 40 cycles and 400 cycles films had thicknesses of $\sim 200$~nm and $\sim 2$~$\mu$m, 
respectively.  Finally, other Rb-M$^{\prime}$-M Prussian blue 
analogue films were investigated, including Rb-Co-Cr, Rb-Cu-Cr, Rb-Zn-Cr, 
Rb-Ni-Fe, Rb-Co-Fe, Rb-Cu-Fe, and Rb-Zn-Fe.\cite{Gardner-thesis,Pajerowski-thesis} 

\begin{figure}[ht]
\includegraphics[width=3.375in]{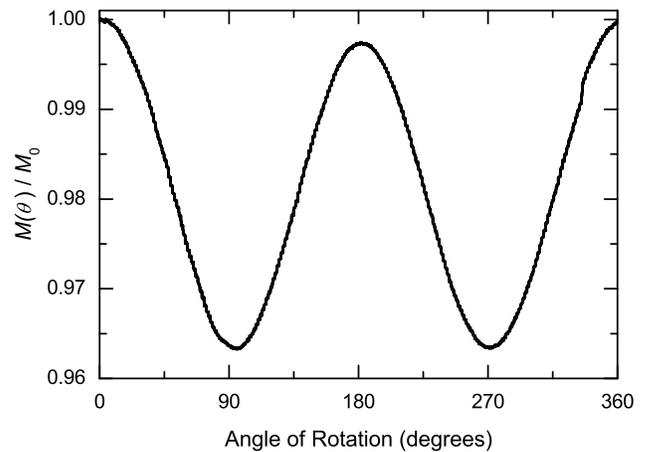}
\caption{The angular variation of $M$ is shown for the case of 
the thick film when $B = 4$ T and $T = 10$~K.  The discrete steps of 
1.5$^{\circ}$ are detectable, and the data were taken continuously at 
each angle that was held for a period of 5~min.  The data 
for the thin film and additional details are available 
elsewhere.\cite{Pajerowski-thesis}}
\end{figure}

The chemical compositions and the physical properties 
of all samples were established by a suite 
of techniques, 
which yielded 
Rb$_{0.7}$Ni$_{4.0}$[Cr(CN)$_6$]$_{2.9} \cdot n$H$_2$O.\cite{Gardner-thesis,Pajerowski-thesis}    
For the magnetization measurements, a commercial (Quantum Design)  
magnetometer was used in conjunction with 
a home-made \emph{in situ} rotator.\cite{dpaj-rotator}  The powder samples were 
mounted in gelcaps, while the film samples were either cut and 
stacked in a plastic box or measured individually in a straw holder.  
A single 400 cycles film, a stack of ten 40 cycles films, and $\sim 100$~$\mu$g of 
powder embedded in eicosane were employed for the cw-ESR measurements performed at either 
50 GHz or 116 GHz, using a resonant 
cavity coupled to a cryostat and superconducting magnet at the 
NHMFL-Tallahassee.\cite{Takahashi}  Transmission and reflection x-ray diffraction  
(XRD) studies were performed at 20 K, 110 K, and 300 K by using 
the instruments at the University of Guelph.  Care was taken to avoid long 
term vacuum pumping of the sample at room temperature, since variations 
due to reversible dehydration-hydration\cite{Ohkoshi,Moritomo2} 
were observed as the (200)
peak shifted to higher $2\theta$ and broadened.  Data were collected for nominally 24 h 
at each temperature, and a blank Melinex film was also measured to 
assist with the background subtraction arising from the solid support.

\begin{figure}[ht]
\includegraphics[width=3.375in]{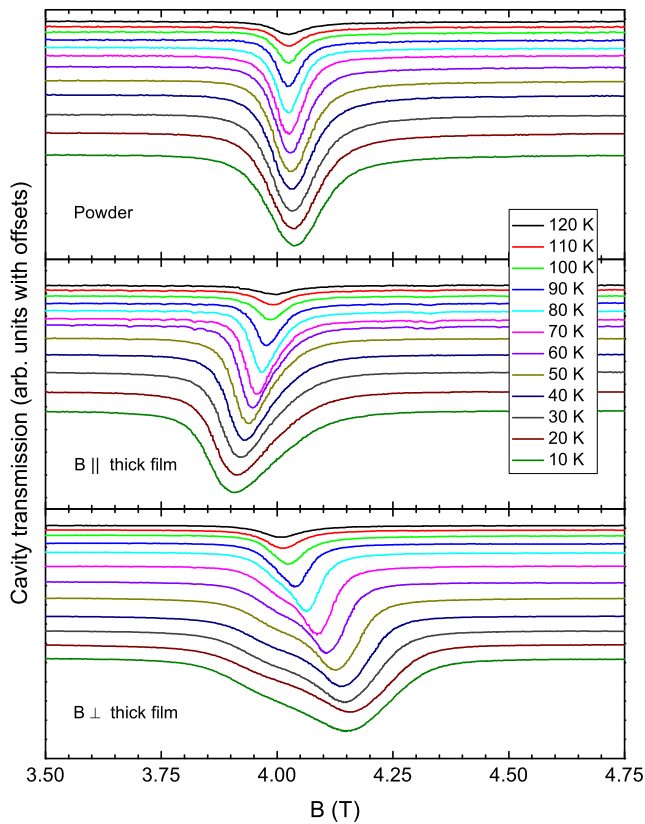}
\caption{(Color online)  The cavity transmission at 116 GHz, as a function of $B$, is shown 
for various temperatures for the powder and the thick (2 $\mu$m) film for 
$B \parallel \mathrm{and} \perp$ to the surface of the film.  The traces are offset 
for clarity.}
\end{figure}

\section{Results}
The anisotropic magnetic response in Prussian blue analogues was initially 
observed in magnetization measurements,\cite{Park1} and this behavior is 
shown for the Rb-Ni-Cr films in Figs.~1 and 2.  
Differences between the ZFC and FC data are related to a spin-glass-like 
response,\cite{Pejakovic,Mydosh} while the anisotropy of the thin and thick films  
is strikingly similar as the external magnetic field is applied 
parallel or perpendicular to the surface of the films, hereafter referred to as 
$B \parallel$ and $B\perp$, respectively.  This behavior is also present, albeit to a somewhat 
weaker degree, in spin-cast samples\cite{Gardner-thesis} but is not observed in 
films that were synthesized in a manner that corrupts their two-dimensional nature by 
generating discontinuities and roughness.\cite{Park2}  

\begin{figure}[ht]
\includegraphics[width=3.375in]{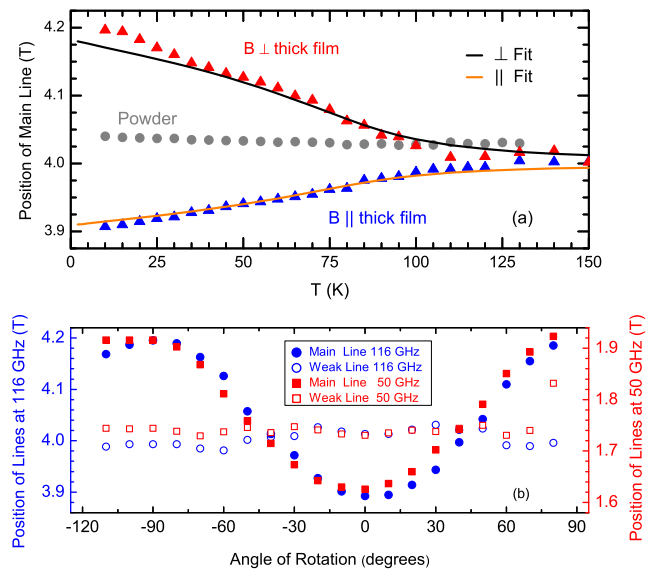}
\caption{(Color online) (a) The temperature dependences of the 
main ESR absorption lines at 116~GHz  
(Fig.~3), for the powder specimen and the thick film with 
$B \parallel \mathrm{and} \perp$ to the surface of the film.  
The solid lines are the results of analysis using 
demagnetization factors, see text.  
(b) Angular dependences of the positions of the main (closed symbols) 
and weak (open symbols) ESR absorption lines at 116 GHz (round symbols and 
left scale) and at 50 GHz (square symbols and right scale).  The 
$B \parallel \mathrm{and} \perp$ orientations are $0^{\circ}$ and 
$\pm 90^{\circ}$, respectively.}
\end{figure}

To date, ESR investigations of Prussian blue analogues have been limited to the 
Rb-Mn-Fe system that ferromagnetically orders near 10~K.\cite{Pregelj,Antal}    
In our work, the nature of the anisotropy was explored, and the 116~GHz 
results for the powder and 2~$\mu$m film are shown in Fig.~3, 
while the data for the 200~nm film 
are consistent with the trends reflected in the thicker film.\cite{Pajerowski-thesis}    
One difference is that the lines of the thin film have a Lorentzian 
shape, while the lines of the thick film have a Gaussian shape, and this 
observation is consistent with the increase of disorder as the films become thicker.    
For the powder, one clear absorption line, with an effective $g = 2.05$, 
is resolved.  The response of the 2 $\mu$m film is similar 
to the powder for $T \gtrsim 100$~K for both orientations of the applied 
magnetic field.  However, 
for $T < 100$ K, the absorption signals are described by two lines, one main line
that is temperature dependent and a weak line that is independent  
of temperature within experimental resolution.  Whereas the main line 
presumably arises from the well coupled Ni$^{2+}$ and Cr$^{3+}$ ions, 
the weak line is associated with trace amounts of 
powder-sized nodules that are observed on the 
surfaces of the films.\cite{Gardner-thesis,Pajerowski-thesis}  

At 10~K, 
the main and weak lines have effective g-values of 2.11 and 2.05 for 
$B \parallel$ and 1.97 and 2.05 for $B \perp$.  The temperature dependences 
of the main line positions are shown in Fig.~4, along with the angular dependences 
of the main and weak lines at 50 GHz and 116 GHz for the 2$\mu$m film at 10 K.  
The angular response of the main line is identical to the behavior observed 
for the magnetization, Fig.~2, and follows a uniaxial $\sin^2(\alpha)$ 
dependence, where $\alpha$ is the angle between $B$ and the surface of 
the film. In addition, the angular dependence of the positions of the main 
lines is the same at both frequencies with a maximum variation of $\Delta B \sim 0.3$~T  
(Fig.~4).  

\begin{figure}[ht]
\includegraphics[width=3.375in]{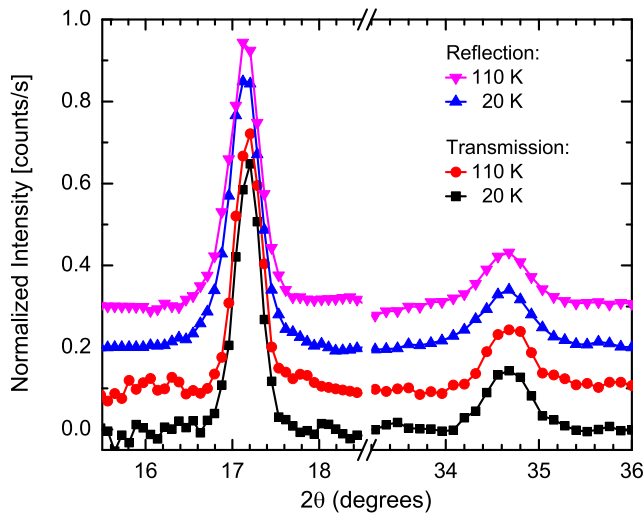}
\caption{(Color online) The XRD pattern collected in reflection and transmission modes 
at 20 K and 110 K.  
The results near the (200) and (400) 
peaks are shown when normalized to the peak values of each data set.  
The data traces are shifted for clarity.  
No changes in the lattice parameters are detected to within 0.005~\AA.}
\end{figure} 

Since deviations from perfect cubic symmetry\cite{Bleuzen} might arise 
when the samples cool through $T_c$, variable 
temperature XRD studies were performed (Fig.~5).  The (200) and (400) peaks at 
17.16$^{\circ}$ and 34.68$^{\circ}$ were monitored in detail, and the results 
do not indicate any change in the lattice parameter through $T_c$, as the 
\emph{Fm}$\overline{3}$\emph{m} (No.~225) cubic symmetry is maintained with a lattice dimension of 
10.33 \AA.  In transmission, the peaks at 24$^{\circ}$ and 30$^{\circ}$ 
were also assignable due to the absence of contributions from the polymer solid support  
in this configuration.

\section{Discussion}
 After inspecting the comprehensive set of experimental results, several points are 
immediately obvious.  Firstly and simply stated, the films possess magnetic  
anisotropy that is not manifested in polycrystalline powder samples that are normally studied.  
Secondly, the  data indicate that the underlying anisotropy prevails for thin and thick films, 
so the anisotropy does not explicitly arise 
from influences coming from direct interaction with the solid support\cite{Gambardella}
but does depend upon the two-dimensional  
organization of the sample generated during the film fabrication process.  Furthermore, the 
values of $T_c$ are independent of the orientation of the magnetic field, 
meaning the anisotropy does not originate from variations of the 
superexchange parameter, $J$.  Thirdly, the ESR results, namely the line shapes of 
the powder spectra and the frequency independence of the magnitude of the line splittings, 
cannot be reconciled by the presence of an anisotropic g-tensor.  Finally, 
magnetostriction or other structural changes are not observed at any temperature, 
so the cubic symmetry is preserved to an extent that does not permit it to be 
a possible explanation of the anisotropy.

With the elimination of several common mechanisms as the possible sources of 
the anisotropic response, magnetostatic interactions remain as a plausible explanation.  
Indeed, the uniaxial nature of the anisotropy is consistent with dipolar 
interactions.  In addition, demagnetizing effects 
($H_{\mathrm{effective}} = H_{\mathrm{lab}} \,-\, N M$, where $N$ is the 
demagnetizing factor) model the data well when using the theoretical value 
to normalize the high field, saturation magnetization value of 
$1.47 \times 10^5$ A/m, namely, 
$\langle S_{\mathrm{Ni}_z} \rangle _{\mathrm{max}} = 1$ and 
$\langle S_{\mathrm{Cr}_z} \rangle _{\mathrm{max}} = 3/2$.\cite{Pajerowski-thesis,Osborn,Vonsovskii}  
The low field magnetization 
in the perpendicular orientation can be reproduced quantitatively from 
the parallel orientation if $N_{\parallel} = 0.07$ and $N_{\perp} = 0.86$, 
Fig. 1a, where domains are expected to obey 
$2N_{\parallel} + N_{\perp} = 1$.\cite{Osborn,Vonsovskii}  
The high field 
magnetization can also be reproduced but not as directly, since the  high field susceptibility 
has a significant experimental uncertainty because $\mathrm{d}M/\mathrm{d}H$ is orders of magnitude 
smaller than $M/H$ in this range.  Nevertheless, the 
uniformly magnetized film limit, \emph{vide infra}, namely $N_{\parallel} = 0$ and 
$N_{\perp} = 1$, reasonably reproduces the observed trends (Fig.~1b).

The ESR data can also be explained by the presence of demagnetization 
effects. \cite{Vonsovskii,Kittel,Kunii}    
Specifically, taking the equations of motion for a spin in $B$ along the z-axis, 
the resonance condition is
\begin{equation}
\omega_0^2=g^2\,\mu_B^2\,[B_z+(N_y-N_z)\mu_o M_z][B_z+(N_x-N_z)\mu_o M_z]\;.
\end{equation}
For a perfect sphere, $N_x = N_y = N_z = 1/3$, 
so the resonance condition should be isotropic
and have no magnetization dependence.  
In practice, there may be small deviations from spherical symmetry for the powder, and
the resonance condition may be written as
\begin{equation}
\omega_{0,\mathrm{powder}}\;=\;g\,\mu_B\,[B- \mu_o\delta M_z] 
\;\;\;,
\end{equation}
where $\delta$ takes care of deviations from spherical symmetry. For the powder data, 
Figs.~2 and 3, a shift of $\sim 10$~mT is present in the fully magnetized state
compared to the paramagnetic state. This observation is consistent with a 
value of $\delta \sim 0.05$, and the shift is similar to the one reported for Rb-Mn-Fe,\cite{Pregelj}  
where it was attributed to demagnetizing effects. 
For a uniformly magnetized film oriented perpendicular 
to $B$, the resonance condition is
\begin{equation}
\omega_{0,\perp}\;=\;g\,\mu_B\,[B- \mu_o M_z]
\;\;\;,
\end{equation}
whereas for the parallel orientation, the resonance condition is
\begin{equation}
\omega_{0,\parallel}\;=\;g\,\mu_B\,[B(B+ \mu_o M_z)]^{1/2}  
\;\;\;.
\end{equation}
Ergo, the temperature dependence of the main lines can be predicted with 
$N_{\parallel} = 0$ and $N_{\perp} = 1$, and the results are in excellent 
agreement with the data (Fig.~4a).

\section{Conclusions}
In summary, the observed magnetic anisotropy of the Rb-Ni-Cr Prussian blue 
analogue films is attributable to demagnetization effects  
arising from their two-dimensional geometry.    
Additional evidence for the magnetic domain-field interactions is 
garnered from the extensive data sets collected on the aforementioned 
Rb-M$^{\prime}$-M Prussian blue analogues.\cite{Gardner-thesis,Pajerowski-thesis}    
Having identified magnetic domains as the origin of the anisotropy, 
additional studies, such as magnetic imaging of the surfaces, will  
provide a deeper understanding of the architecture and dynamics of the domains.  
Finally, a systematic approach for determining the nature of magnetic anisotropy 
in coordination polymers has been presented and will be important as this class 
of materials is investigated for physical properties applicable to spintronic 
applications.

\acknowledgments
We acknowledge conversations with M.~F.~Dumont, M.~W.~Lufaso, 
and A.~Ozarowski.   
This work was supported, in part, by NSERC, CFI, and NSF through DMR-0804408 (SH), 
DMR-1005581 (DRT), DMR-0701400 (MWM), and the NHMFL via cooperative agreement 
under NSF DMR-0654118 and the State of Florida.  
We thank Ben Pletcher and 
the Major Analytical Instrumentation Center (MAIC), Department 
of Materials Science and Engineering, University of Florida, 
for help with the EDS, SEM, and TEM work.

\end{document}